\documentclass[pra,aps,twocolumn,amsmath,amssymb]{revtex4}

\usepackage{graphicx}% Include figure files
\usepackage{dcolumn}% Align table columns on decimal point
\usepackage{bm}% bold math
\usepackage[extension=xxx]{hyperref}

\newcommand{\beq}{\begin{equation}}
\newcommand{\eeq}{\end{equation}}
\newcommand{\beqa}{\begin{eqnarray}}
\newcommand{\eeqa}{\end{eqnarray}}

\def\beq{\begin{equation}}

\usepackage{color}

\begin{document}

\title{Ultrarobust calibration of an optical lattice depth based on a phase shift}
\date{\today}

\author{C. Cabrera-Guti\'errez$^1$, E. Michon$^1$, V. Brunaud$^1$, T. Kawalec$^2$, A. Fortun$^1$, M. Arnal$^1$, J. Billy$^1$, D. Gu\'ery-Odelin$^1$}

\affiliation{$^1$ Universit\'e de Toulouse ; UPS ; Laboratoire Collisions Agr\'egats R\'eactivit\'e, IRSAMC ; F-31062 Toulouse, France} 
\affiliation{CNRS ; UMR 5589 ; F-31062 Toulouse, France}
\affiliation{$^2$ Marian Smoluchowski Institute of Physics, Jagiellonian University, \L{}ojasiewicza 11, PL-30348 Krak\'ow, Poland}

\begin{abstract}
We report on a new method to calibrate the depth of an optical lattice. It consists in triggering the intrasite dipole mode of the cloud by a sudden phase shift. The corresponding oscillatory motion is directly related to the intraband frequencies on a large range of lattice depths. Remarkably, for a moderate displacement, a single frequency dominates this oscillation for the zeroth and first order interference pattern observed after a sufficiently long time-of-flight. The method is robust against atom-atom interactions and the exact value of the extra external confinement of the initial trapping potential. 
\end{abstract}
\maketitle

Optical lattices play a key role in quantum simulations performed with cold atoms \cite{bloch2012,bloch2008}. 
They are produced by the interference of far-off resonance laser beams and the resulting spatially modulated intensity defines  
purely conservative and defect-free potentials in which atoms evolve. By an appropriate set of lasers interfering at various angles, any lattice geometry can be generated \cite{grynberg,Greiner2001,RMPOberthaler,Becker2010,Gyu2011,Tarruell2012}. The band structure and topology can also be directly engineered by tuning the phase, intensity or polarization of the interfering laser beams.

An important issue for experiments is to know, with a high accuracy, the potential really experienced by the atoms. For instance, the tunneling rate depends exponentially on the potential depth for a sufficiently large depth. A large error on this quantity can therefore be made with a calibration method not precise enough. In principle, the lattice depth can be readily calculated once the characteristics of the laser beams (waist, detuning, polarization and power) and of the atomic transition (atomic polarizability and saturation intensity) are known. However, the precise knowledge of the quantities that characterize the beam inside the vacuum chamber, where atoms are manipulated, turns out to be difficult to establish with an accuracy better than 10-20 \%. Indeed, the beam propagates through the windows of the vacuum chamber yielding possible distorsions of the beam profile combined with potential birefringence effects. The laser beam may not have a $M^2$ factor larger than 1, etc ... %The laser beam may also differ from the perfect Gaussian beam, which would need to be taken into account in calculations.% it is difficult to include in the calculations the details of the laser beam cross section, which may differ from the perfect Gaussian beam

To calibrate in situ the lattice depth, different methods have been developed including parametric heating \cite{Friebel1998}, Rabi oscillations \cite{Ovchinnikov1999}, Raman-Nath diffraction \cite{Gould}, expansion from a lattice \cite{Cristiani} to cite a few.  Each method has a certain range of validity, and one shall be careful in the error bar estimations.  The most accurate methods rely on the analysis of Bose Einstein condensate diffraction pattern. %However, in order to get a reliable estimate of the lattice depth through this method, one should be careful to minimize the role of the mean-field and of the extra confinement usually superimposed to the lattice, which could modify the diffraction pattern observed.

In this article, we propose a new calibration method that has the advantage of being independent of the interaction strength, the quantum statistics, and is valid over a very large range of potential depths. This method simply consists in triggering, by a sudden phase shift, the intrasite dipole oscillations of the gas inside the optical lattice, and to relate the frequency of the observed oscillations to the appropriate interband transition. 

The paper is organized as follows. In Sec.~\ref{sec1}, we discuss the calibration method based on the diffraction pattern observed after exposing the quantum gas to a pulsed optical lattice. Section \ref{sec2} addresses the technique which consists in analyzing the interference pattern observed once the lattice is suddenly switched off. Our new method is detailed in Sec.~\ref{sec3}. We then conclude the paper with some general comments on the method.
 
Our experiment produces nearly pure rubidium-87 BEC of $10^5$ atoms in the lowest hyperfine level $F=1,m_F=-1$. It relies on a hybrid trap made of a quadrupole magnetic field and a crossed dipole trap \cite{PRL2016}.  %of maximum gradient 1.8 T/m
The 1D optical lattice is created by superimposing two mutually coherent and far-off resonance counter-propagating laser beams at $\lambda=$1064 nm. The corresponding optical potential reads: 
\begin{equation}
V_L(x)=-U_0\cos^2\left(  \frac{\pi x}{d} + \theta_0 \right) 
\end{equation}
and has a lattice spacing $d=\lambda/2=532$ nm. With our notations, a change of the phase $\theta_0$ by $90^o$ moves the minima to the positions of the maxima. In the following, the depth $U_0$ of the lattice is measured in units of the the lattice characteristic energy, $E_L=h^2/(2md^2)$ \cite{footnote1}: $U_0=sE_L$. In our experiment, the 1D optical lattice is superimposed to the horizontal optical guide of the hybrid trap \cite{PRL2016}. 

The relative phase, $\theta_0$, between the two beams generating the optical lattice is controlled in time via synthesizers whose frequencies are imprinted on light with AOMs (acousto-optic modulators). The phase $\theta_0$ can therefore be modified on a very short timescale (on the order of a few nanoseconds) or can be modulated in time. This phase-modulation technique is quite standard and was recently used in quantum simulation with cold atoms in lattices to engineer the tunneling rate \cite{ChuPRL,ArimondoPRL,Oberthaler,PRXJean,Hofstadter,Haldane,sengstock,chin,reitter,goerg}.

\section{Matter wave exposed to a pulsed optical lattice}
\label{sec1}

\begin{figure}[t]
\centering
\includegraphics[width=8cm]{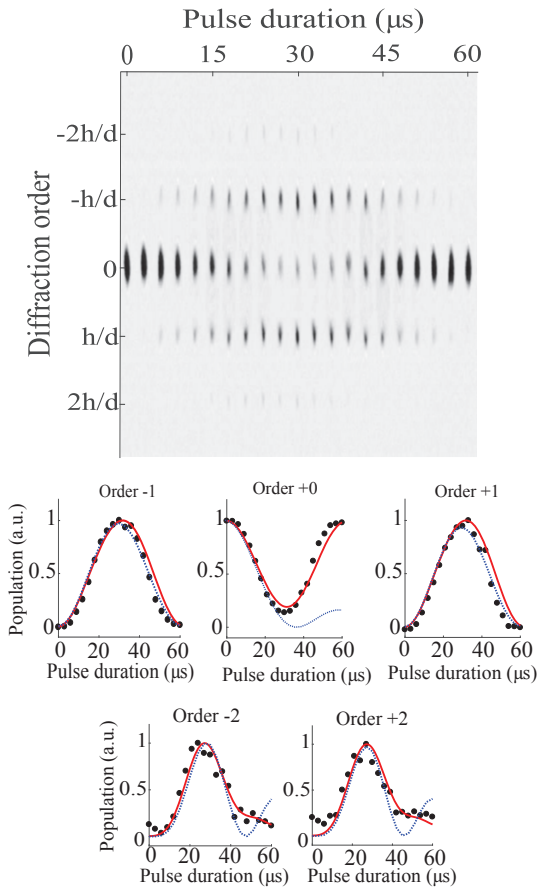}
\caption{(a) Diffraction of a rubidium-87 Bose Einstein condensate from a pulsed one-dimensional optical lattice for $s\simeq 2.7$. The different images correspond to different pulse durations separated by 3 $\mu$s. (b) Orders -2, -1, 0, 1 and 2: Experimental data (dark points), Raman-Nath approximation (blue dotted lines), Mathieu functions analysis (red solid lines).}
\label{figure1}
\end{figure}

The most common method used to calibrate the depth of an optical lattice consists in shining the light that generates the lattice for a short amount of time, $\tau$, and to analyze the corresponding diffraction pattern observed after a sufficiently long time-of-flight. Examples of such diffraction patterns are shown in Figure~\ref{figure1}. 

\subsection{Analytical solution}

The quantitative analysis of the diffraction patterns is based on the formalism of Mathieu functions \cite{McLachlan,horne}. Indeed, the stationary solutions $\Phi(x)$ of the one-dimensional Schr\"odinger equation associated with the periodic potential $V_L(x)$ obeys the Mathieu's equation:
\begin{equation}
\frac{\textrm{d}^2\Phi}{\textrm{d}\tilde{x}^2}+[a+2q\cos(2\tilde{x})]\Phi=0,
\label{eqmathieu}
\end{equation}
with $\tilde{x}=\pi x/d$, $a=4E/E_L+2s$ and $q=s$. The solutions of Eq.~(\ref{eqmathieu}) are the Mathieu's functions $M(a,q;\tilde{x})$. According to Bloch-Floquet's theorem, those solutions can be recast in the form $M(a,q;\tilde{x})=\textrm{e}^{ir\tilde{x}}f(a,q,\tilde{x})$ for fixed values of ($a$, $q$) where $r$ is an integer referred to as the characteristic exponent, and $f$ a periodic complex valued function having the same spatial period as that of the lattice. The Mathieu's functions have the same parity as their characteristic exponent. For fixed values of $q$ and $r$, only one value of $a$ provides a solution of Eq.~(\ref{eqmathieu}). This specific value, $a_r$, is referred to as the characteristic value. 

The BEC has a size much larger than the spatial period $d$; we can therefore assume that the initial wave function is constant. We choose a normalization to unity for one spatial period. The initial wave function therefore reads, $\psi(\tilde x,0)=(2\pi)^{-1/2}$. To derive explicitly the diffraction pattern, we expand the initial wave function of the BEC on the Mathieu's basis:
\begin{eqnarray}
|\psi(0)\rangle &= & \sum\limits_{r=0}^{\infty} \langle M(a_r,q)| \psi(0) \rangle | M(a_r,q)\rangle
 \nonumber \\
 & = & \sum\limits_{j=0}^{\infty}\frac{c_{2j}}{\sqrt{\pi}}|M(a_{2j},q)\rangle,
\end{eqnarray}
with $c_{2j}=(\sqrt{2}/\pi)\int_{0}^{\pi}M(a_{2j},q;\tilde{x})\textrm{d}\tilde{x}$. During the time over which the pulse is applied, each component $2j$ evolves freely according to its energy $E_{2j}/E_L=a_{2j}/4-q/2$: 
\begin{equation}
\psi(\tilde{x},\tau)=\textrm{e}^{-iq\frac{E_L\tau}{2\hbar}}\sum\limits_{j=0}^{\infty}\frac{c_{2j}}{\sqrt{\pi}}M(a_{2j},q;\tilde{x})\textrm{e}^{-ia_{2j}\frac{E_L\tau}{4\hbar}}.
\end{equation}
From this expression, we infer the population of the n$^{th}$ diffraction order:
\begin{equation}
P_n(U_0,\tau)=\left|\int_{0}^{2\pi}\frac{1}{\sqrt{2\pi}}\psi(\tilde{x},\tau)\textrm{e}^{-2i n\tilde{x}}\textrm{d}\tilde{x}\right|^2.
\end{equation}

\subsection{Raman Nath approximation}

The previous expression can be simplified in the limit of very short pulses. Indeed, in this case, atoms have not enough time to change their velocity during the pulse and the kinetic energy can be neglected \cite{Gould}. This approximation is commonly called the Raman Nath approximation. More qualitatively, this criterium reads $\tau \ll 2\pi/\omega_0$ with $\omega_0=(2\pi^2U_0/md^2)^{1/2}=(E_L/\hbar)s^{1/2}$ where the angular frequency $\omega_0$ is associated with the harmonic approximation of the minima of $V_L(x)$. Note that this criterium is all the more restrictive at low lattice depths. In practice, the Raman Nath approximation allows to calibrate accurately lattice with a sufficiently large depths $s$ ($s>5$ typically).

Under this approximation, the periodic potential is directly imprinted on the phase of the wave function:
\begin{eqnarray}
\psi(x,\tau) & = & \textrm{e}^{-iV_L(x)\tau/\hbar}\psi(x,0), \nonumber \\
\psi(x,\tau) & \propto & \sum\limits_{n=-\infty}^{+\infty}i^nJ_n\left(\frac{U_0\tau}{2\hbar}\right)\textrm{e}^{ink_Lx}\psi(x,0),
\label{diffracramannath}
\end{eqnarray}
yielding
\begin{equation}
P_n(U_0,\tau) \propto \left|  J_n\left(\frac{U_0\tau}{2\hbar}\right)  \right|^2.
\label{eqbessel}
\end{equation}

In Figure \ref{figure1}, we have represented a set of data at the breakdown edge of the Raman Nath approximation. Fitting the data with the Bessel functions for the orders -2,-1,0,1,2 of the diffraction pattern (see Eq.~(\ref{eqbessel})) yields a depth of $2.78 \pm 0.52\; E_L$. If we apply the Mathieu functions to the different orders, we get a much more accurate estimation $2.68 \pm 0.01\;E_L$. In Fig.~\ref{figure1}, we clearly see that this latter fit gives a much better account of the zeroth order time-evolution and of the wings observed on the orders $\pm 2$. The accuracy for this method in the low depths regime therefore comes at the expense of a detailed fit procedure over many orders and involving the Mathieu functions. The breakdown of the Raman-Nath approximation also generates an envelope of the time-evolving diffraction pattern that is shaped by caustics \cite{LaburtheTolra}.

\section{Sudden expansion method}
\label{sec2}

Another calibration method consists in studying the interference pattern obtained after a sudden switch-off of the lattice and a subsequent long time-of-flight. For this purpose, one shall first load the BEC adiabatically in the optical lattice. 

\begin{figure}[t]
\centering
\includegraphics[width=8cm]{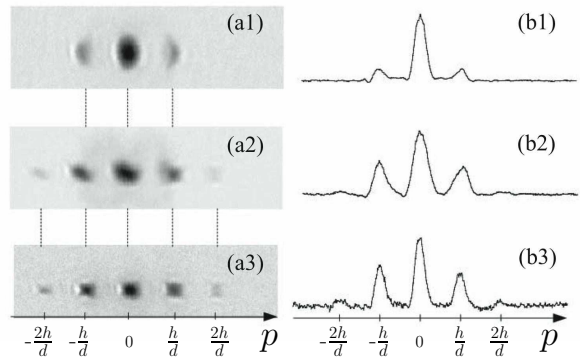}
\caption{(a) Time-of-flight pictures (expansion during 25 ms) obtained after a sudden release of an adiabatically loaded Bose Einstein condensate into an optical lattice for different depths $U_0=$1.9, 6.6, 11.4 $E_L$. (b) Integrated profile corresponding to each image. }
\label{figure2}
\end{figure}

Once the BEC has been produced in our hybrid trap, the optical lattice is ramped up adiabatically in 30 ms using a S-shape variation in time of the intensity of the lattice beams. The time duration for the ramp  is large compared to $\hbar/\mu$ where $\mu$ is the BEC chemical potential \cite{AdCr}. As a result, the BEC reaches a steady state for which it is splitted in small BECs occupying the different lattice sites. The lattice is then abruptly switched off. The wave packets, initially located at the lattice sites, then expand freely and interfere with one another yielding a serie of regularly spaced peaks. Figure \ref{figure2} provides three examples of such interference patterns for different lattice depths. The resulting spatial interference pattern is then analyzed to extract the value of the depth of the potential $U_0 $\cite{Cristiani}.

Assuming that the size of the condensate is large compared to the lattice spacing, we can assume that the wave function is periodic and reads 
\begin{equation}
\psi(x)=\sum\limits_{m=1}^{N_s}f_0(x-md)\textrm{e}^{i\varphi_m},
\label{3.17}
\end{equation}
where $N_s$ is the number of sites, $f_0$ corresponds to the wave function in a single lattice site and $\varphi_m$ accounts for the relative phase in the different lattice sites. For large lattice depths ($s>15$), the tunneling between adjacent sites becomes negligible and the BEC in each site will be rapidly out of phase \cite{dalibardstock}. Indeed, the phase is proportional to the chemical potential of the clouds in each lattice site which may differ from site to site due to the loading of a different number of atoms in the lattice sites.

An absorption image is then taken after a sufficiently long time-of-flight and reveals the initial density in momentum space
\begin{equation}
\begin{aligned}
\left|\tilde{\psi}(p)\right|^2&=\left|\tilde{f_0}(p)\right|^2\left|\sum\limits_{m=1}^{N_s}\textrm{e}^{i\varphi_m}\textrm{e}^{-ipmd/\hbar}\right|^2,
\end{aligned}
\label{trfourier}
\end{equation}
where $\tilde{f_0}(p)$ is the Fourier transform of $f_0(x)$. The expression of the second factor depends on the phase relation between the sites \cite{dalibardstock}. For low depth, the tunneling favors a uniform phase and we can assume $\varphi_m=\varphi_0$, and we get
\begin{equation}
\left|\textrm{e}^{i\varphi_0}\sum\limits_{m=1}^{N_s}\textrm{e}^{-ipmd/\hbar}\right|^2 =\frac{\sin^2\left(\frac{N_spd}{2\hbar}\right)}{\sin^2\left(\frac{pd}{2\hbar}\right)}.
\end{equation}
To get the population in each order, we need a model function for $f_0$. As a first example, we assume a Gaussian function corresponding to the wave function of the harmonic oscillator obtained by expanding $V_L$ about its minima \cite{pedri}. This approximation is expected to be all the more valid as the magnitude of $s$ is large. The Fourier transform then reads $\tilde{f_0}(p) \propto \exp(-p^2a_0^2/2\hbar^2)$ where $a_0=(\hbar/m\omega_0)^{1/2}$ is the harmonic oscillator length associated with $\omega_0$. The population $\Pi_n$ for the order $n$ is then readily obtained from Eq.~(\ref{trfourier}) with $p_n= nh/d$:
\begin{equation}
\Pi_n=\Pi_0\textrm{e}^{-(2\pi na_0/d)^2}=\Pi_0\textrm{e}^{-2n^2/s^{1/2}}.
\label{eq3.21}
\end{equation}
Under this Gaussian approximation the lattice depth can therefore be directly inferred from (\ref{eq3.21})
\begin{equation}
s_{gauss}=\left(\frac{2n^2}{\textrm{ln}(P_n)}\right)^2,
\label{eq3.22}
\end{equation}
where $P_n=\Pi_n/\Pi_0$
However, for relatively shallow lattice ($s<15$), this approximation is not accurate since the tunneling favors a spreading of the wave function. A correction was numerically studied in Ref.~\cite{pedri}, and an analytical formula proposed in Ref.~\cite{Cristiani} for $s<5$, 
\begin{equation}
s \simeq s_{gauss}P_n^{-1/4}.
\label{eqw}
\end{equation}
There is no available interpolation formula at intermediate depths ($5<s<15$). In this range of parameters, it is therefore difficult to infer with accuracy the correspondance between the population in the interference orders and the depth. 
%Interestingly and according to our 1D numerical simulations, the population in the different orders is marginally affected by the interactions. 
In the following section, we propose another strategy to deduce the lattice depth accurately on a broad range of lattice depths ($0.5 < s < 50$). 

\section{The sudden phase shift method}
\label{sec3}

\begin{figure}[t]
\centering
\includegraphics[width=8cm]{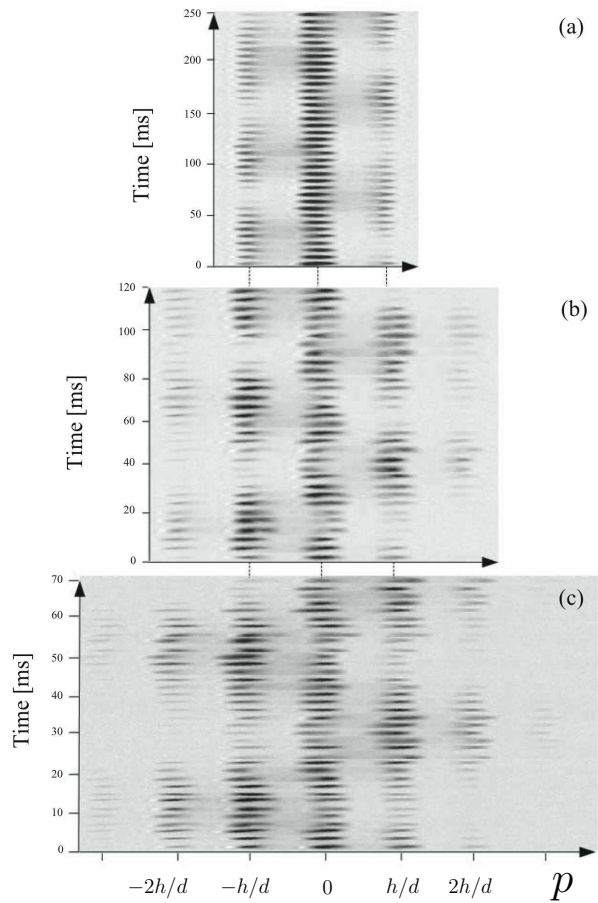}
\caption{Oscillation in momentum space (obtained after a 25 ms time-of-flight) triggered by an initial phase shift $\theta_0=25^o$ and for different depths $U_0=$1.9 (a), 6.66 (b), 11.4 $E_L$ (c).}
\label{figure3}
\end{figure}

\begin{figure*}[!t]
   \begin{center}
\includegraphics[width=17cm]{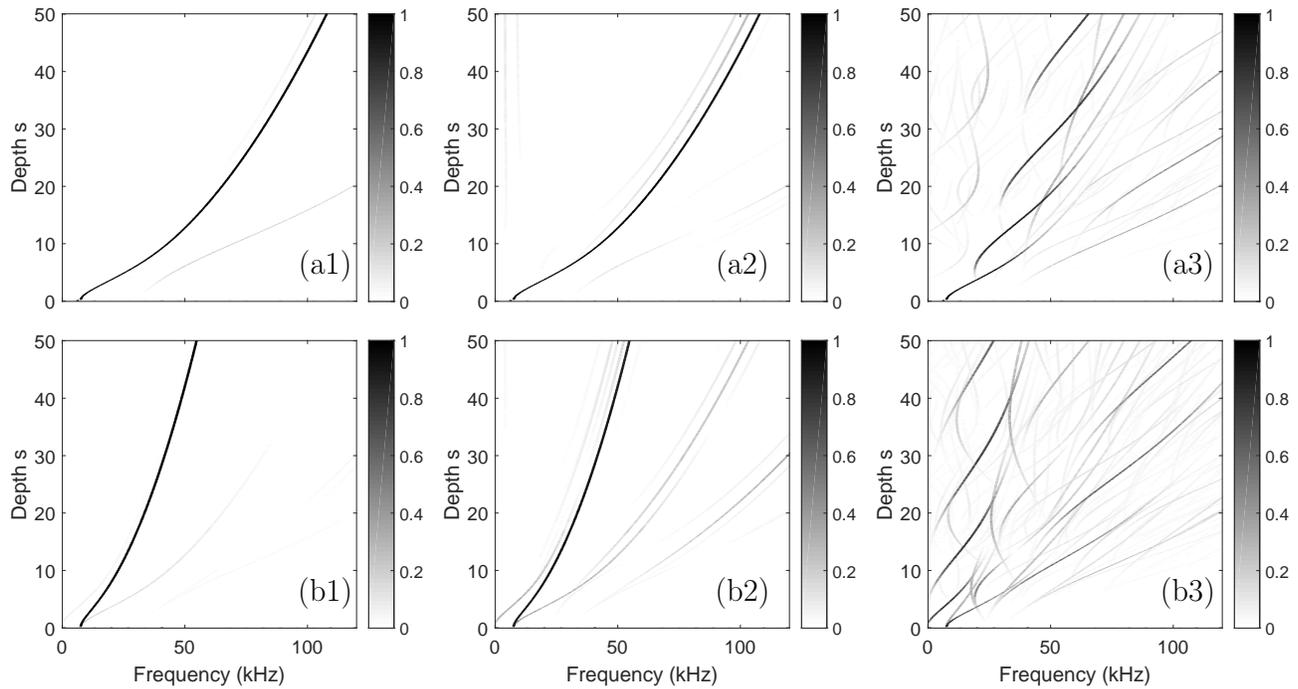}
\end{center}
\caption{Populations $\Pi_n(\nu)$ in momentum space versus frequency $\nu$, for the orders $n=0$ (a) and $n=\pm 1$ (b). The populations $\Pi_n(\nu)$ are defined as the Fourier transform of the oscillation pattern resulting from the initial quench. The quench corresponds to an offset of the initial potential by an angle $\theta_0=10^o$ ($a_1$ and $b_1$), $\theta_0=25^o$ ($a_2$ and $b_2$) and $\theta_0=70^o$ ($a_3$ and $b_3$). The gray scale corresponds to a decimal scale.}
\label{figure4}
\end{figure*}

For the new method that we propose, we start as the previous method: a BEC is first adiabatically loaded into an optical lattice. 
We then perform a sudden phase shift of the optical lattice, $\theta(0^+)=\theta_0 < 90^o$. The analysis of the subsequent motion of the wave function can be readily obtained by expanding the initial wave function onto the Bloch states with zero pseudo-momentum:
\begin{equation}
\Psi(x,t=0)=\sum_{n=1}^{+\infty} c_{n, 0}u_{n, 0}(x),
\end{equation}
where the functions $u_{n,q}(x)$ are the Bloch functions for a given depth $U_0$ associated with the band $n$ and the pseudo-momentum $q$ such that $-\pi/d \le q \le \pi/d$. The time evolution reads
\begin{equation}
\Psi(x,t)=\sum_{n=1}^{+\infty} c_{n,0}e^{-i\frac{E_{n}(q=0) t}{\hbar}}u_{n,0}(x).
\label{eqevolutionintime}
\end{equation}
where $E_{n}(q)$ are the Bloch energies of the $n^{th}$ band for a lattice of depth $U_0$.
The evolution is observed after a long time-of-flight and for different holding time, $t$, in the lattice after the phase shift. Figure \ref{figure3} provides three examples of such series of snapshots for different depths where the back and forth oscillation in the lattice potential wells can be clearly seen. The pattern observed in those images is nothing but the modulus of the Fourier transform of the wave function: 
\begin{equation}
P(p,t)=|\tilde  \Psi (p,t)|^2 \propto  \left|\int  \Psi(x,t)e^{ipx/\hbar}dx\right|^2.
\label{tfpsi}
\end{equation}
To characterize the oscillatory behavior observed in momentum space, we take the time Fourier transform of $P(k,t)$ for the different orders and introduce the population for each order $n$:
\begin{equation}
\Pi_n(\nu) \propto \int P\left(\frac{nh}{d},t\right)e^{i\nu t} dt.
\end{equation}
In Figure~\ref{figure4}, we plot  $\Pi_{0}(\nu)$ and $\Pi_{\pm 1}(\nu)$ for different lattice depths $s$, and initial offset angle $\theta_0$. The main result of this article concerns the zeroth order for relatively small values of $\theta_0$ (see Fig.~\ref{figure4} for $\theta_0=10^o$). Indeed, only one frequency essentially appears for a very broad range of potential depths from $s=0.5$ to 50. The one to one correspondance between the frequency and the depth provides an interesting calibration method valid over a very wide range of lattice depths. Furthermore, for low depth, the first order of the oscillating pattern in Fourier space has even a better contrast than the zeroth order, and therefore constitutes a reliable alternative to fit the data.

\begin{figure}[t]
   \begin{center}
\includegraphics[width=8cm]{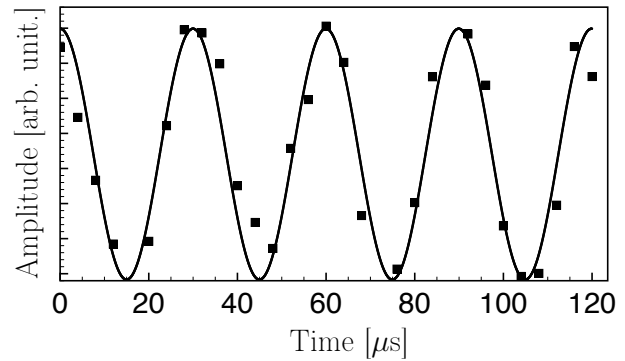}
\end{center}
\caption{Population of the zeroth order of the diffraction pattern of Fig.~\ref{figure3}b as a function of time: Experimental data (black squares), Fit (solid line). We find a frequency $\nu=33.3$ kHz corresponding to a depth $U_0=6.66 E_L$.}
\label{FigureDemo}
\end{figure}

One may wonder why only one frequency dominates the zeroth order population $\Pi_{0}(\nu)$ for relatively low offset angle $\theta_0$. Two main effects are responsible for this interesting behavior: first the small initial phase shift dramatically limits the number of Bloch states over which the initial state is projected (to typically 4), and second, the integral over space in Eq.~(\ref{tfpsi}) selects only the contribution of states that have an even parity when $k=0$. As a result the contribution of second and fourth Bloch bands are washed out. The frequency of oscillation of the zeroth order population $\Pi_{0}(\nu)$ is therefore given by 
\begin{equation}
\nu_0=\frac{E_3(q=0)-E_1(q=0)}{h}.
\end{equation}
In Figure \ref{FigureDemo}, we provide an analysis based on this relation between the depth and the measured frequency associated to the diffraction pattern of Fig.~\ref{figure3}b. In this case, we find a lattice depth equal to $6.66E_L$.

Similarly, at low offset angle, the first order population $\Pi_{+1}(\nu)$ is dominated by one frequency $\nu_1$: %where $E_n(q)$ are the Bloch energies of the $n^{th}$ band for an lattice of depth $U_0$. %A similar argument provides the single frequency, $\nu_1$, observed in $\Pi_{+1}(\nu)$ 
 \begin{equation}
\nu_1=\frac{E_2(q=0)-E_1(q=0)}{h}.
\end{equation} 
In this case, the small offset angle still limits the number of Bloch states on which the initial state is projected. However, no selection rule applies in this case, which explain the higher number of frequencies for the first order population than for the zeroth order (compare Figs.~\ref{figure4}a2 and b2).

\begin{figure}[t]
\centering
\includegraphics[width=8cm]{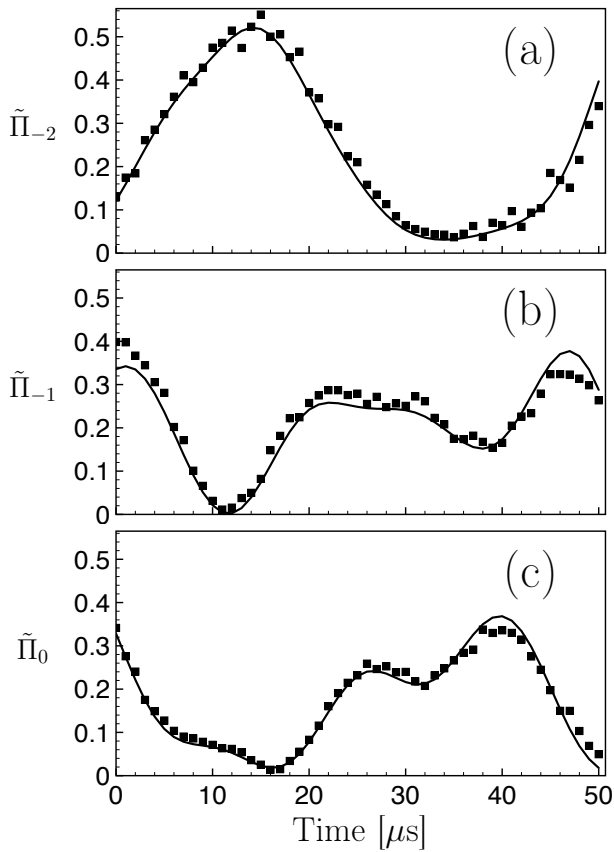}
\caption{Normalized populations $\tilde \Pi_m$ in the m$^{th}$ order of the diffraction pattern as a function of time for $\theta_0=70^o$: (a) $m=-2$, (b) $m=-1$ and (c) m=0. The depth of the optical potential is $U_0=9.4E_L$.}
\label{figure4plus}
\end{figure}

Interestingly, for a larger initial angle $\theta_0=25^o$, the dominant frequency in the spectrum of $\Pi_0$ and $\Pi_{+1}$ remains unchanged (see Fig.~\ref{figure4}). Our calibration method is therefore quite robust against the value of the initial angle. For sake of comparison, we have also plotted similar spectra for a very large initial offset angle $\theta_0=70^o$. As intuitively expected, for such a large initial angle the initial wave function is projected on many bands and many interband frequencies are involved in the subsequent oscillatory motion. In Figure \ref{figure4plus}, we have plotted the normalized populations 
\begin{equation}
\tilde \Pi_n (t)= \frac{P(nh/d,t)}{\sum_n P(nh/d,t)},
\end{equation}
 in the m$^{th}$ order (for $m=0,-1$ and $-2$) of the diffraction pattern as a function of time for $\theta_0=70^o$ along with the prediction obtained from Bloch analysis. We find a perfect agreement between experiment and theory in this very far-off-equilibrium dynamics without any adjustment. This confirms that our analysis is valid and robust even outside the single frequency regime.  

A major advantage of our calibration method is that it is immune to both atom-atom interactions and the specific value of the extra external confinement provided by the initial trap. Note that even if the loading is not perfectly adiabatic, the method still holds. This result has been proved with extensive numerical simulation of the Gross-Pitaevskii equation\cite{PRL2016}. The independence of the frequencies $\nu_0$ and $\nu_1$ with respect to interactions results from the fact that the motion triggered by the sudden initial phase shift corresponds to the so-called dipole mode. It corresponds to an oscillation of the intrasite center of mass which is not affected by the strength of the interactions.

\section{Conclusion}

The dipole mode in ordinary traps, also referred to as the Kohn's mode, characterizes the center of mass oscillation, and is consequently not affected by two-body interactions, the temperature nor the statistics. This is the reason why it is used to experimentally calibrate the value of the trap frequencies. In this article, we have shown its interest for in situ determination of lattice depths. For this purpose, we apply a sudden change in position of the lattice, small compared to the lattice spacing period (less than 100 nm), and observe the subsequent motion of the wave packets after a time-of-flight. Besides the well known advantages of dipole oscillations mentioned above, this method provides a simple mapping between the single frequency observed in the motion and the lattice depths, and turns out to be robust against the precise value of the position shift because of selection rule for interband transition.

This work was supported by Programme Investissements d'Avenir under the program ANR-11-IDEX-0002-02, reference ANR-10-LABX-0037-NEXT, and the research funding grant ANR-17-CE30-0024-01. M.~A. acknowledges support from the DGA (Direction G\'en\'erale de l'Armement).


\begin{thebibliography}{99}
\bibitem{footnote1}
The characteristic energy $E_L=4E_{\rm rec}$ where $E_{\rm rec}=\hbar^2k^2/2m$ with $k=\frac{2\pi}{\lambda}$ is the recoil energy associated with the absorption of one photon of one of the laser that generates the lattice beam.
\bibitem{AdCr}
Strictly speaking, the adiabatic loading of the lattice is even more difficult to reach as the lattice depth is large since the tunneling rate decreases exponentially with the lattice depth.
\end{thebibliography}
\end{document}